\documentclass[twocolumn,letterpaper]{IEEEAerospaceCLS}

\usepackage[]{graphicx}    
\usepackage{amsfonts}
\usepackage{amsmath}
\usepackage{lipsum}
\usepackage[utf8]{inputenc}
\usepackage{booktabs}
\usepackage{makecell}
\usepackage{url}
\usepackage{multicol}
\usepackage{dblfloatfix}

\newcommand{\ignore}[1]{}

\usepackage{xcolor}

\graphicspath{{./figures/}}

\begin{document}
\title{Topological Simplification of Signals for Inference and Approximate Reconstruction}

\author{%
	Gary Koplik \\
	Geometric Data Analytics
	\and 
	Nathan Borggren
	\and Sam Voisin \\
	Geometric Data Analytics \\
	\and Gabrielle Angeloro \\
	Geometric Data Analytics \\
	\and Jay Hineman \\
	Geometric Data Analytics \\
	\and Tessa Johnson \\
	Geometric Data Analytics \\
	\and
	Paul Bendich \\
	Geometric Data Analytics \\
	Department of Mathematics \\
	Duke University
}

\maketitle

\thispagestyle{plain}
\pagestyle{plain}

\begin{abstract}
As Internet of Things (IoT) devices become both cheaper and more powerful, researchers are increasingly finding solutions to their scientific curiosities both financially and computationally feasible. When operating with restricted power or communications budgets, however, devices can only send highly-compressed data. Such circumstances are common for devices placed away from electric grids that can only communicate via satellite, a situation particularly plausible for environmental sensor networks. These restrictions can be further complicated by potential variability in the communications budget, for example a solar-powered device needing to expend less energy when transmitting data on a cloudy day. We propose a novel, topology-based, lossy compression method well-equipped for these restrictive yet variable circumstances. This technique, \textit{Topological Signal Compression}, allows sending compressed signals that utilize the entirety of a variable communications budget. To demonstrate our algorithm's capabilities, we perform entropy calculations as well as a classification exercise on increasingly topologically simplified signals from the Free-Spoken Digit Dataset and explore the stability of the resulting performance against common baselines.
\end{abstract}


\section{Introduction}
\label{sec:intro}

Time series arise whenever numerical values are collected over time.
Time series classification involves training a model on a set of labeled time series signals, and then using the model to predict
those labels on a test set. Applications of time series classification abound: for example, using signals of labeled EEG time series to predict \cite{Chaovalitwongse2006eeg} epileptic brain activity.
Multiple methods \cite{Bagnall2017ts} for time series classification exist in the literature, including a variety \cite{Fawaz2017dts} of deep learning techniques.

This paper explores a novel lossy compression technique, \emph{Topological Signal Compression} (TSC).
Figures \ref{fig:compression_example} and \ref{fig:tsc_simplification_real_data} demonstrate this technique.
Illustrative results are shown using the Free-Spoken Digit Dataset (FSDD)\cite{fsdd}.
The key findings of this paper appear in Figures \ref{fig:ml_results}, \ref{fig:compressed_confus}, and \ref{fig:entropy}, which show that TSC both preserves information content and can maintain classifier performance while significantly reducing the size of signal needed to achieve said performance. Furthermore, TSC achieves this capability in an interpretable way at arbitrarily high compression levels and with compression on the margin having only a local effect on the reconstructed signal.

\subsection{Motivation}
\label{sec:Mot}

We are motivated by the following abstraction of a common paradigm: we imagine that the time series signals to be classified are collected by any one of potentially many edge devices, and that the classification itself must happen at a central device.\footnote{Although there are plenty of cases where one would be inclined to resolve classification on the edge as opposed to at a central device, this is not always possible, for example, if a task required information from \textit{multiple} devices before classification. One may not even be interested in classification at all, instead focusing on returning as much relevant raw signal data as possible.}
In addition to classification accuracy, we will also judge success based on the amount of transmission between the edge devices and the central device. 

Examples of this paradigm include: a) the Internet-of-things (IoT), where on-device power constraints or a low-bandwidth communications network can preclude the transmission of full signals to the central node; b) surveillance applications, where excessive transmission between a drone and a central computer increases the chances of counter-surveillance measures detecting and thus disrupting the classification process. 
In general, we call these \emph{constrained communications} (CC) scenarios.

The Topological Signal Compression (TSC) algorithm proposed in this paper serves as a generic lossy compression step useful in any time series compression and / or classification task that must take place under constrained communications. 
TSC is adaptable to levels of CC, as the algorithm permits transmission at exactly the level of transmission permitted by the scenario.
Furthermore, TSC is stable to changing levels of CC. On a theoretical level, TSC performs a localized and thus more stable compression when removing more points---if one were to remove an additional point from a signal already compressed via TSC, then the resulting signal reconstruction would only be affected around the removed point. This stability is further demonstrated experimentally in this paper with the graceful degradations shown in both classifier performance and entropy levels as the CC level increases as well as with a Dynamic Time Warping (DTW) analysis comparing compressed then reconstructed signals to the original signals.

We contrasted TSC with several additional lossy compression methodologies in this paper---the Opus codec, the Discrete Fourier Transform (DFT), and Piecewise Aggregate Approximation (PAA). Opus is an audio-specific codec whereas TSC generalizes to compressing any time series data. Given our choice of an audio dataset tailored to the narrow strengths of Opus for classification tasks, we found in our experimental analysis that Opus better-maintains performance over increased compression levels and noise than TSC. Unlike Opus, however, TSC is also flexible to any precise compression cutoff, whereas Opus is harder to use to compress a signal to a specific byte size. Furthermore, when using Opus in practice on the Free-Spoken Digit Dataset, we were unable to compress beyond roughly 90\% of the original size of a signal, whereas TSC can be run at arbitrarily high compression percentages. Finally, in our machine learning exercises, we found only TSC and Opus maintained both accuracy and stability with respect to higher levels of compression as well as when noise was added to the data. TSC was thus the only algorithm considered in this paper that had the union of generalizability, flexibility, and stability that we believe to be important in a highly-compressed data transmission scenario with a variable communications budget.

\subsection{Compression With a Variable Communications Budget}

Consider a sensor network where a given edge device can send no more than a small but ever-changing quantity of bytes of collected information to the central device at a moment in time. As a more concrete example, consider the use case of IoT over the ocean in the \textit{Ocean of Things (OoT)} project \cite{waterston2019ocean}.  Since the raw data itself is of high value on this project, one would not want to only send summary statistics or any sort of classification results alone, as the time series data itself is of great value. For example, sending high-resolution spatiotemporal ocean environmental data would be valuable to oceanographers in evaluating and improving ocean models. Additionally, for floats to be able to send data from anywhere in the ocean, the only viable means of data transmission is via satellite, which drastically shrinks the maximum possible communication. budget to only a few hundred bytes.\footnote{Iridium Short Burst Data, for example, has a transmission budget constraint of 340 bytes \cite{iridium}.}

Although there may be a fixed bandwidth constraint, the possibility of floats clustering would require at times having a more restrictive communications budgets for each float.\footnote{A single satellite can only process so many messages at one time. Thus, if 1000 floats are in a cluster trying to report to one satellite at one moment in time, they will likely strain the communications channel at that moment, even if each float is transmitting within its original communications budget.} Moreover, since these floats are capable of collecting and reporting a range of multimodal data products of variable relevance for different use cases, the triage to prioritize which data to send will lead to situations where even for a fixed communications budget, there will be a variable \textit{remaining} budget to send information from a given modality.

TSC addresses these circumstances for the OoT use case. For any spare space in a given communications budget that would be wasted otherwise, one could simply run TSC to return more points from a signal that use the exact number of remaining available bytes, a task that cannot easily be achieved by Opus or Piecewise Aggregate Approximation. As for triaging which data to send, TSC could update dynamically to requests by a human or an automated anomaly detection algorithm to prioritize sending more of one signal at the expense of another. Since TSC generalizes to any modality of signal data, one can simply revise each modality's byte constraint and then run TSC for all the modalities. One could even factor in environmental constraints. For example, if devices were solar-powered, and battery levels were low on a cloudy day, signals could be compressed more than normal to save power by transmitting fewer bytes, preventing the devices from temporarily running out of battery power. Though one would have the same flexibility with the Discrete Fourier Transform, the instability of the compressed data, particularly at higher levels of compression, would make inference between signals variably-compressed with DFT difficult.

\subsection{Outline}

The rest of this paper proceeds as follows.
Persistent homology and its use in Topological Signal Compression is discussed in Section \ref{sec:tsc}.
Then Section \ref{sec:data} introduces the real dataset used for illustration, with machine-learning and entropy experiments described in Section \ref{sec:ml}. 
The paper discusses how TSC can be used in practice relative to several competing compression methodologies in Section \ref{sec:discussion}, and concludes in Section \ref{sec:conclusion}.

\section{Topological Signal Compression}
\label{sec:tsc}

The typical mathematical model of a one-dimensional signal is a real-valued function $f$ on a closed interval $[a,b]$, 
but for this work we imagine that the interval has been sampled at a discrete set of time points $a = t_0 < t_1 < t_2 < \ldots < t_n = b$,
and so $f$ is given by its values at each of these time points.
This section outlines the \emph{persistence diagram} summary of $f$, and then
describes our proposed scheme for using persistence diagrams to transmit parsimonious approximations of $f$.

\subsection{Persistent Homology}
\label{sec:ph}

\begin{figure}[h!]
	\includegraphics[width=3.4in]{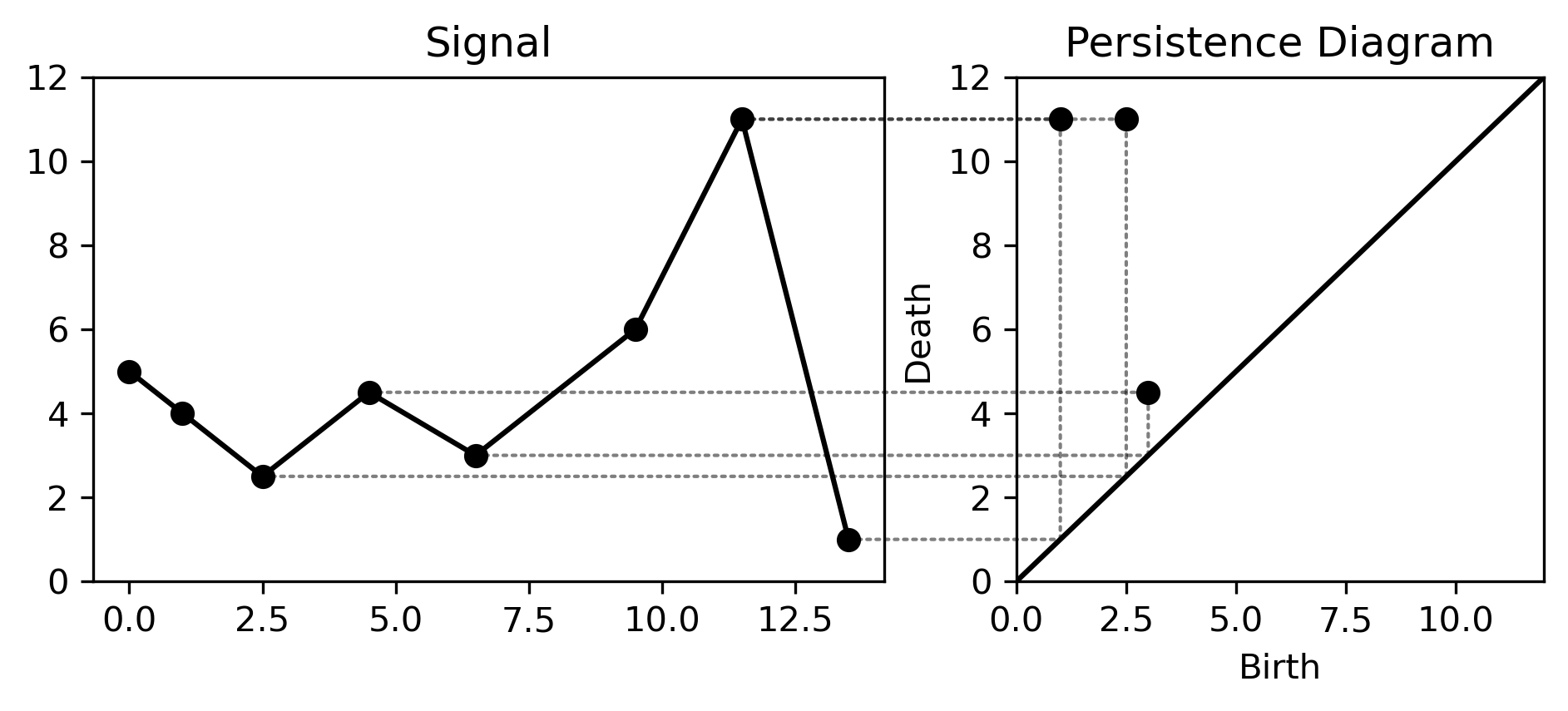}    
	\caption{On the left a signal with 8 points is shown. The Persistence Diagram is shown on the right and formed by sweeping upwards. Components are born at local minima and die at local maxima, destroying the more recently born component. We follow the convention of pairing the global minimum and maximum.}
	\label{fig:signal_pers}
\end{figure}

The signal $f$ is summarized by its \emph{zero-dimensional persistence diagram} (\cite{edelsbrunner2000topological} \cite{Chazal2009proximity}) $D_0(f)$. Intuitively, one need only understand the following in the context of a signal. Here, persistent homology tracks connected components as we sweep a horizontal line vertically from negative infinity to positive infinity. Components are \textit{born} at local minima, and \textit{die} at local maxima, destroying the more recently born component of the two components merging.
The diagram $D_0(f)$ plots the births and deaths of components as dots in the plane. The vertical distance of a dot above the 45 degree line birth = death represents the \textit{persistence} of a component within the filtration. See Figure \ref{fig:signal_pers} for an example of a signal and its corresponding persistence diagram.

Persistent homology thus gives us an \textit{ordering} on our connected components. Dots on the persistence diagram that are close to the diagonal die soon after being born. By allowing us to identify low-persistence components, persistent homology shows us the parts of the signal most likely corresponding to noise.
More precisely, persistence diagrams enjoy a \emph{stability theorem} (\cite{CohenSteiner2007},\cite{Chazal2009proximity}) that states, roughly, that diagrams corresponding to functions which are small perturbations of each other will differ mostly by the presence or absence of low-persistence dots.  We note that some results (e.g, \cite{Bendich2016brains}) have shown these low-persistence dots to still have classification power in machine learning applications. 

\subsection{Topological Simplification}
\label{sec:tsp}

\begin{figure}[h!]
	\includegraphics[width=3.4in]{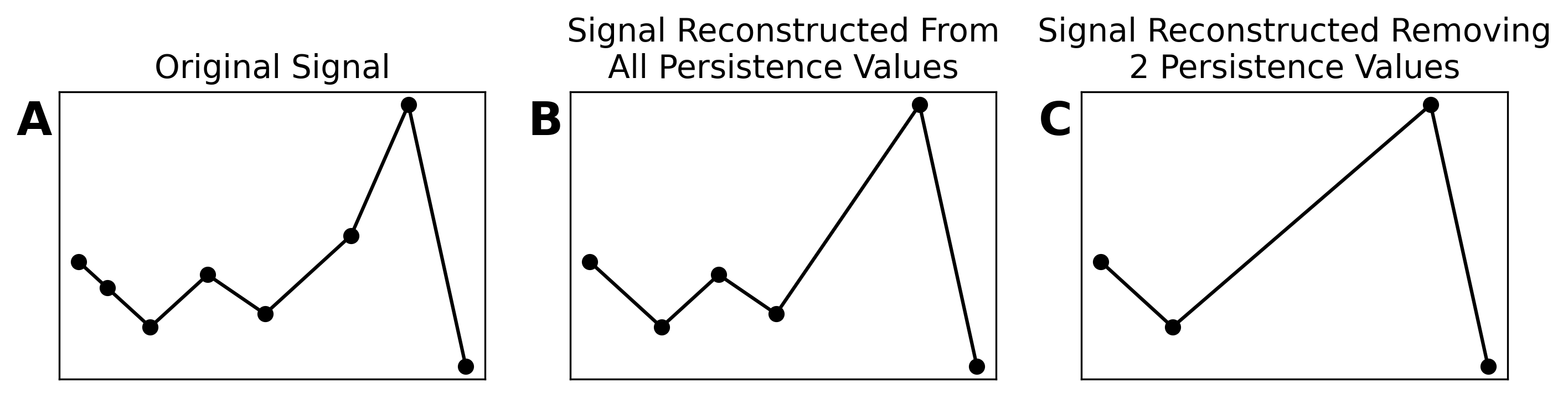}    
	\caption{An example of running our topological simplification algorithm. (A) is our original signal, borrowed from Figure \ref{fig:signal_pers}. (B) is the baseline topologically simplified compression, which keeps only critical points, thus dropping the two non-critical points in the signal. (C) drops the two further points corresponding to the smallest persistence value on the persistence diagram, which can be validated comparing against Figure \ref{fig:signal_pers}.}
	\label{fig:compression_example}
\end{figure}

\begin{figure}[h!]
	\centering
	\includegraphics[width=3.4in]{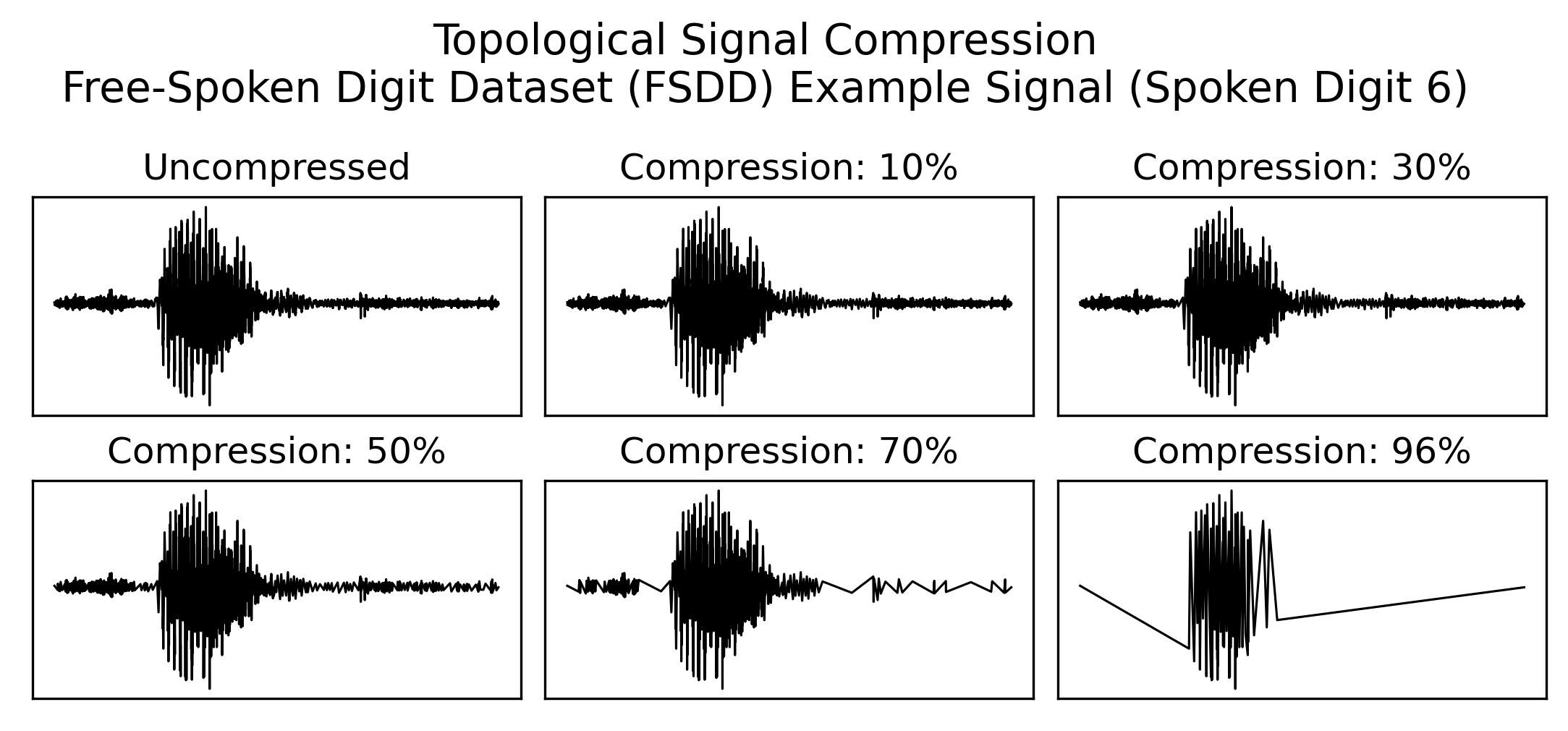}
	\caption{Example from the Free-Spoken Digit Dataset (FSDD) of a person speaking the integer 6, shown with increasing levels of topological simplification.}
	\label{fig:tsc_simplification_real_data}
\end{figure}

Inspired by the typical situation where dots of low-persistence correspond to noise in the signal, 
our simplification method enables the reconstruction of the signal as it would be without the values of least persistence, thus keeping the more prominent features of the signal at the expense of noise. Figures \ref{fig:compression_example} and \ref{fig:tsc_simplification_real_data}
show examples of this simplification on synthetic and real data, respectively. 

The theoretical force behind the algorithm is the \emph{Morse Cancellation Lemma} (MCL, \cite{morse_theory}, and also see \cite{Laudenbach2013} for a self-contained proof). In the simplest version which is all that we require here, the MCL states that, if ${\bf u} = (b,d)$ is 
the dot of lowest persistence in the diagram $D_0(f)$ of a one-dimensional signal $f$, then there exists a signal $g$ defined on the same domain whose persistence diagram $D_0(g)$ is exactly the same as $D_0(f)$ except that ${\bf u}$ has been removed.
For example, panel C of Figure \ref{fig:compression_example} shows one such $g$ corresponding to the signal $f$ in panel B of the same figure.

In effect, $g$ is formed from $f$ by ``un-kinking'' the pair of critical points of lowest persistence.
The reason that this action does not cause global change in the signal stems from the MCL\footnote{Technically, this lemma requires that the function values of the two neighbors of any given critical point be distinct. If this assumption fails, we could create non-unique solutions; however, if we sort first by persistence and then, for example, by index value of the points as input into the algorithm, the output will still be both correct and consistent, even though it may not be unique.}, which guarantees that the dot of lowest persistence corresponds to a pair of \textit{horizontally adjacent} critical points in the time domain, thus guaranteeing that un-kinking that pair of critical points will not un-kink any other pair.
The MCL can of course be applied iteratively, as demonstrated in Figure \ref{fig:tsc_simplification_real_data}.

Bauer et al. \cite{bauer2010total} have noted a relationship between persistence-based simplification in one dimension and total variation-based denoising. Simplification via the MCL extends as well to functions defined on two-dimensional domains (Edelsbrunner et al \cite{edelsbrunner2006persistence} give an algorithm
for this two-dimensional simplification), but there are theoretical issues with higher-dimensional domains.
More relevant to the current work, the reader may have noted that there are in fact infinitely many functions $g$ whose persistence diagrams 
are the required simplifications of $D_0(f)$. For example, the long diagonal line in panel $C$ of Figure \ref{fig:compression_example} could be replaced by any monotonically increasing function defined on the same sub-interval without affecting the persistence diagram. 
Several works address ways to choose the ``right'' type of $g$; for example, Poulenard et al \cite{Poulenard2018tfo} give a general technique 
for finding a $g$ that satisfies a user-specific cost function. 
The perspective we take in this paper is that we simply transmit the $(t,f(t))$ values needed for the central device to make this choice. Note that all illustrations and machine learning tasks throughout this paper perform piecewise linear interpolation when reconstructing signals.

\section{Data}
\label{sec:data}
We used the Free-Spoken Digit Dataset (FSDD) created by Zohar Jackson, C\'esar Souza, Jason Flaks, Yuxin Pan, Hereman Nicolas, and Adhish Thite \cite{fsdd}. This audio dataset was produced by six male speakers and consists of recordings of spoken digits 0 through 9. The 3000 examples were recorded at 8kHz, with each person recording 50 samples of each digit. Finally, the samples are trimmed to have minimal silence at both the beginning and end of each audio clip.

\section{Machine Learning and Entropy}
\label{sec:ml}

\begin{figure}[h!]
	\centering
	\includegraphics[width=2.in]{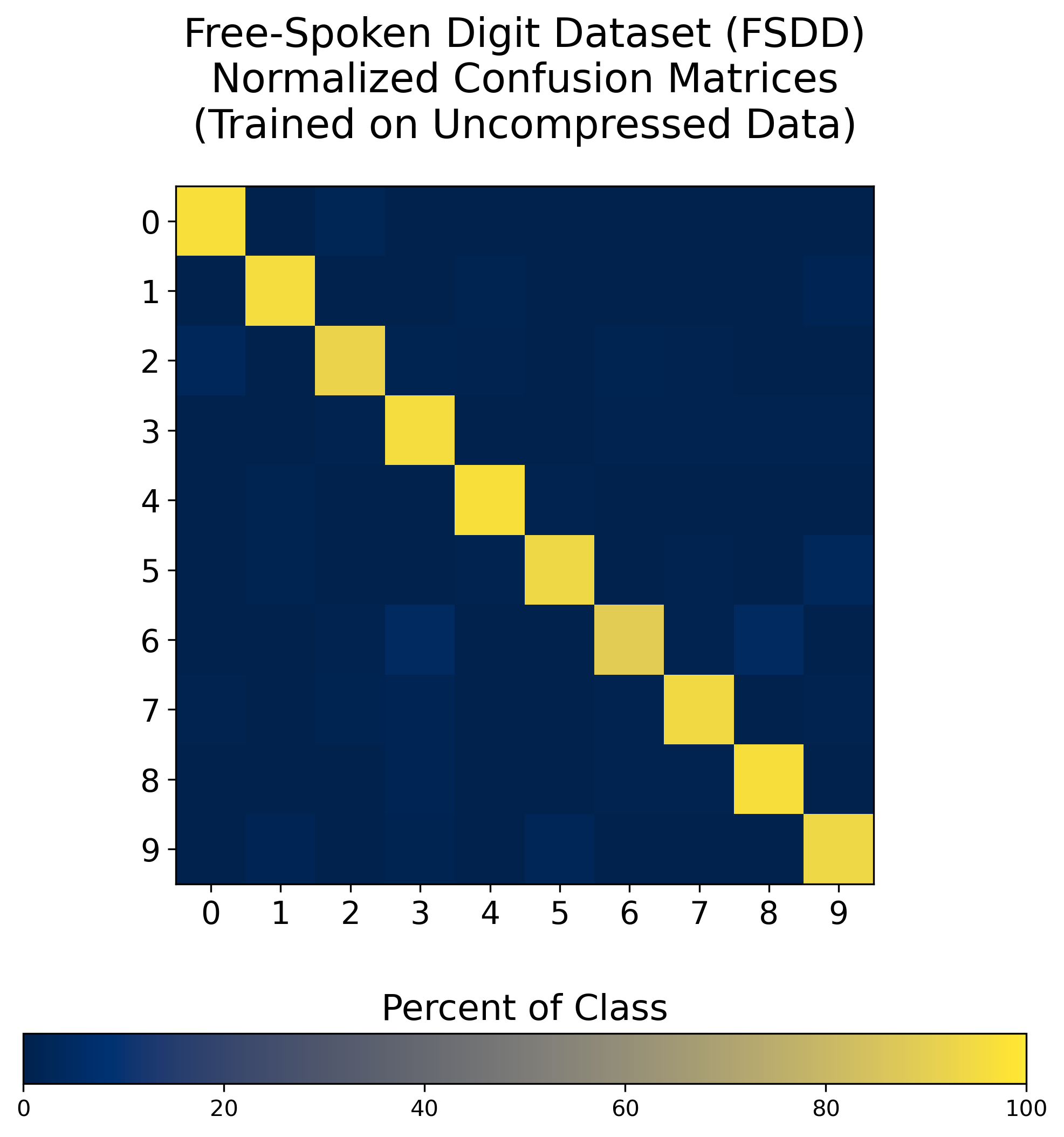}
	\caption{Confusion matrix for classification of Free-Spoken Digit Dataset (FSDD) digits with no compression of the data before featurization.}
	\label{fig:uncompressed_confus}
\end{figure}

For our baseline machine learning exercise, we classified FSDD spoken digits using Mel-Frequency Cepstrum Coefficient (MFCC) featurizations run through a Convolutional Neural Network (CNN) with 5-fold cross-validation. The MFCC featurization performs a short-time Fourier analysis with a binning scheme motivated by the anatomy of the human ear. For more on MFCC, see \cite{bogert1963quefrency} and \cite{logan2000mel}. By this process, each sample was transformed into a 20 x 16 feature vector. For this initial exercise, we achieved a mean cross-validated accuracy of approximately 97\%. The confusion matrix for these baseline results is reported in Figure \ref{fig:uncompressed_confus}.

\begin{figure*}[h!]
	\centering
	\includegraphics[width=7in]{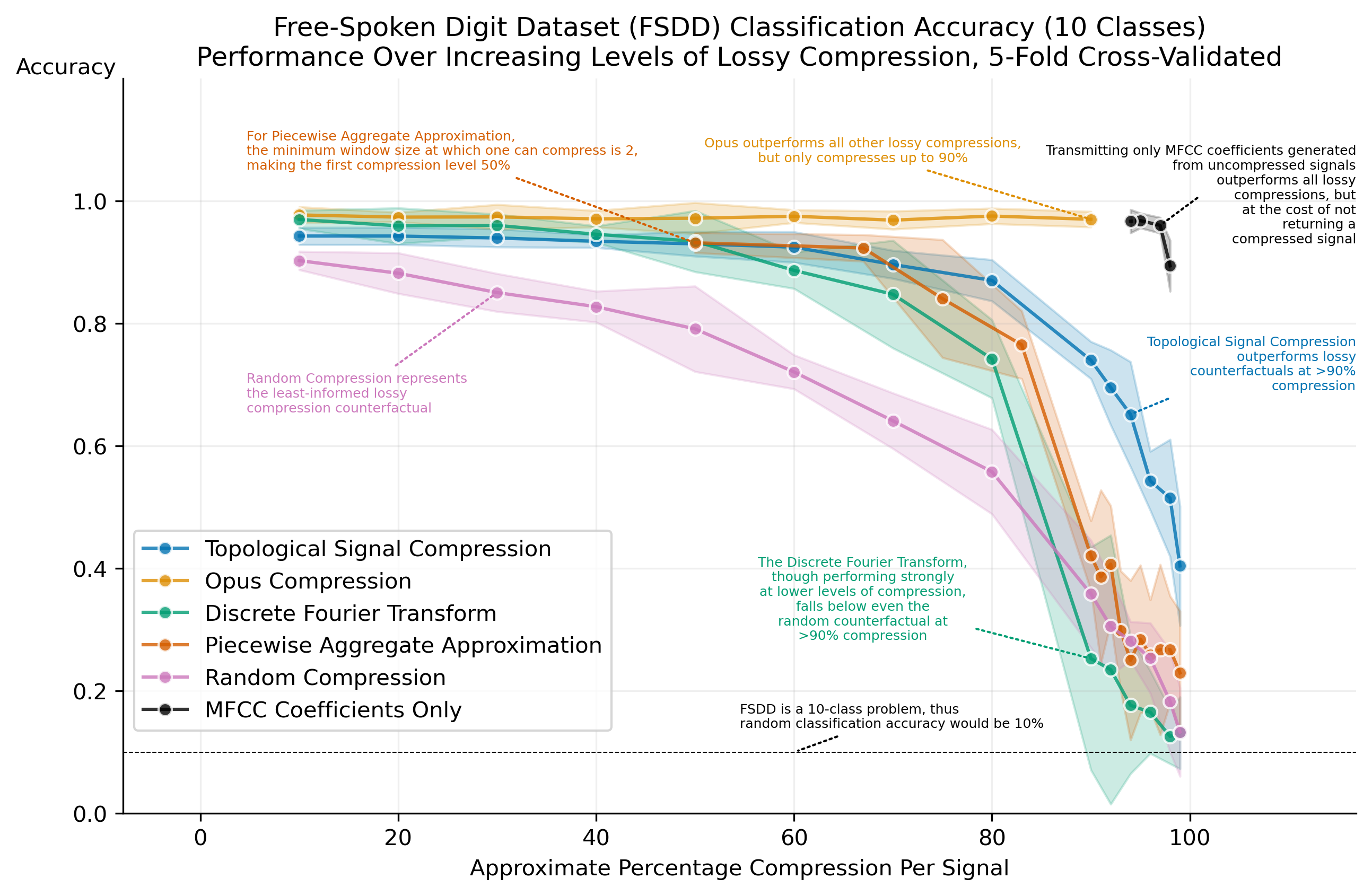}
	\caption{Accuracy for 10-label classification task with Convolutional Neural Network using MFCC featurizations generated from increasingly compressed signals. Error bars represent 2 times the standard deviation of accuracy over the 5-fold cross-validated results at each compression level.}
	\label{fig:ml_results}
\end{figure*}

\begin{figure}
	\centering
	\includegraphics[width=3.4in]{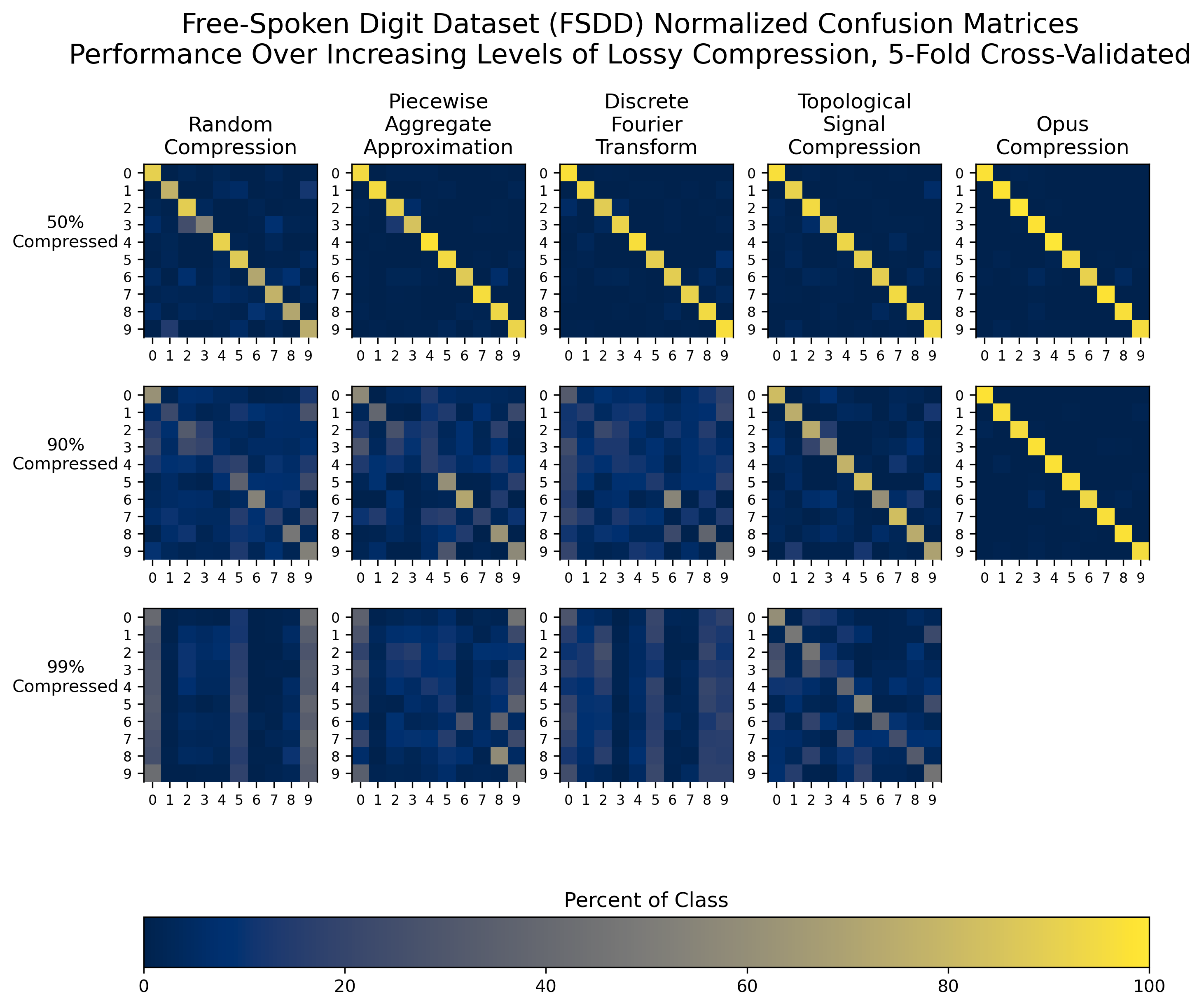}
	\caption{Confusion matrix for classification of 10 digits of selected compression percentages for our 5 compression methodologies.}
	\label{fig:compressed_confus}
\end{figure}

\begin{figure*}
	\centering
	\includegraphics[width=7in]{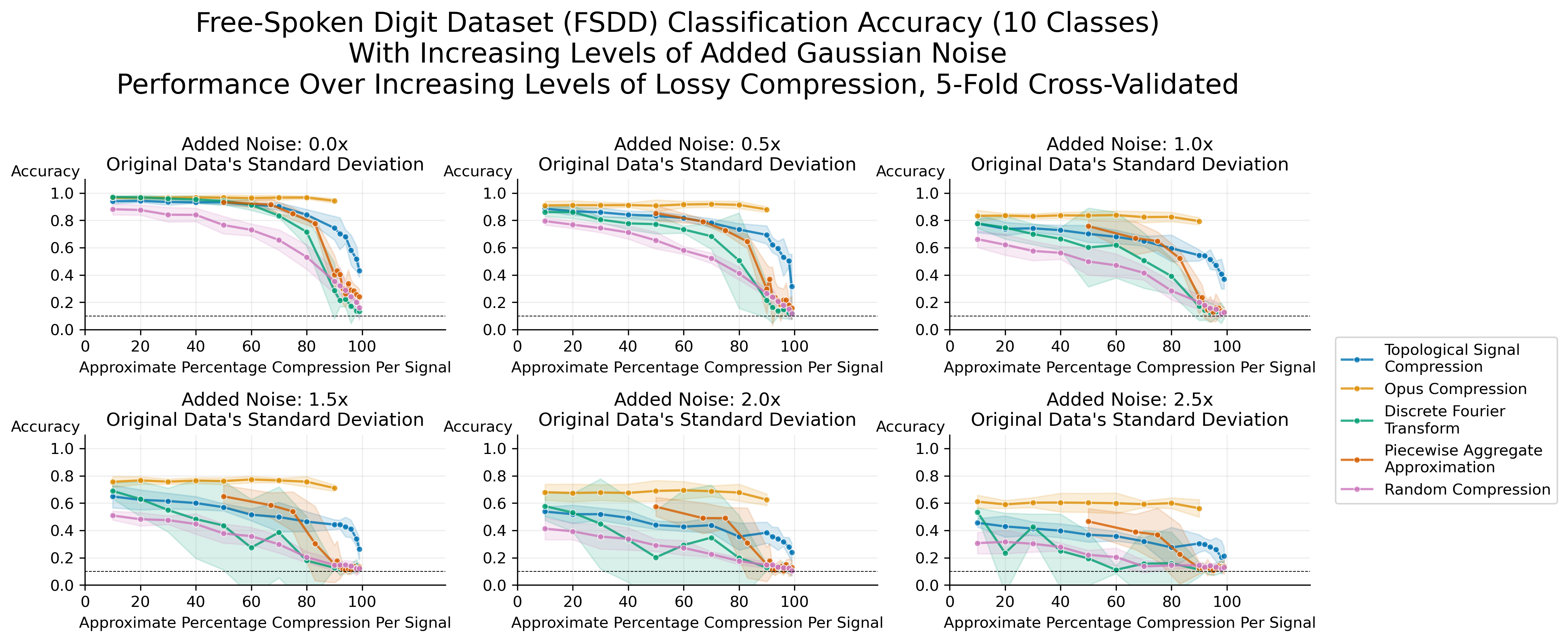}
	\caption{Accuracy of various compression methodologies over increasingly noisy FSDD dataset. Gaussian noise as high as 2.5 times the noise in the standardized signals was added to the dataset. Accuracy declines overall as noise increases, but \textit{relative} performance between compression schemes is mostly consistent with Figure \ref{fig:ml_results}, with one major exception being DFT compression performing noticeably worse with more noise. Although TSC's performance suffers more than Opus with increasing levels of noise, TSC still outperforms the other compression methodologies at greater than 90\% compression. Error bars represent 2 times the standard deviation of accuracy over the 5-fold cross-validated results at each compression level.}
	\label{fig:accuracy_with_noise}
\end{figure*}

We then repeated this cross-validated classification pipeline, only we built the MFCC features with increasingly topologically simplified signals from FSDD. We demonstrate an example of various levels of topological simplification on an example from FSDD in Figure \ref{fig:tsc_simplification_real_data}.

In addition to TSC, we considered four competing forms of lossy compression to act as counterfactuals. We first considered Opus, a codec designed exclusively for lossy audio compression \cite{opus}. We then compressed signals using the Discrete Fourier Transform (DFT) by representing a signal with its Fourier coefficients and then sending only a subset of those coefficients to reconstruct the signal \cite{schafer2012sfa}. Next, we compressed signals using Piecewise Aggregate Approximation (PAA), which uniformly partitions a signal using a fixed window size and returns the mean function value within each window \cite{keogh2001dimensionality}. Finally, as the most naive control, we considered ``Random Compression,'' where a random subset of $(t, f(t))$ pairs are returned as a lossy compression. For a comparison of how these competing compression methods affect FSDD data relative to TSC as shown in Figure \ref{fig:tsc_simplification_real_data}, see Figure \ref{fig:other_compressions_real_data}.

Our cross-validated accuracy results are reported in Figure \ref{fig:ml_results}. As a reference point for if we were solely concerned with machine learning accuracy as opposed to the ability to reconstruct a signal, we also included ``Uncompressed'' results in black, which considers the ``compression'' to be sending only MFCC featurizations, where we varied the number of MFCC coefficients returned. As expected, when we sacrifice returning any sort of signal to only return features, we are able to maximize machine learning performance at higher compression levels.

As for the lossy compression schemes considered in Figure \ref{fig:ml_results}, we see that Opus dominates in machine learning accuracy over increasing levels of compression, but this success requires a couple of qualifications. First, given the sole focus of Opus on audio compression, it is unsurprisingly the most successful compression method of the five methods considered here given FSDD is an audio dataset. The other four compression methods considered, including TSC, are agnostic to the type of signal data. Second, using the minimum level of bitrate compression possible, Opus was unable to achieve an average compression on FSDD data greater than 90\%, thus limiting its potential ability to be used in highly constrained communications scenarios.

Once we reach a compression of greater than 90\%, at which point Opus is no longer an option for this dataset, TSC-compressed data maintains a mean cross-validated classification accuracy of roughly 15-25 percentage points better than the competing compression methods.

We then explored accuracy for the compression methods on each of the 10 labels in FSDD. The confusion matrices in Figure \ref{fig:compressed_confus} illustrate that machine learning on TSC-compressed data better maintains within-label classification accuracy at 90\%+ compression relative to the non-Opus compression methods, as demonstrated by the more pronounced diagonal structure in the TSC confusion matrices.

\begin{figure}[h!]
	\centering
	\includegraphics[width=3.4in]{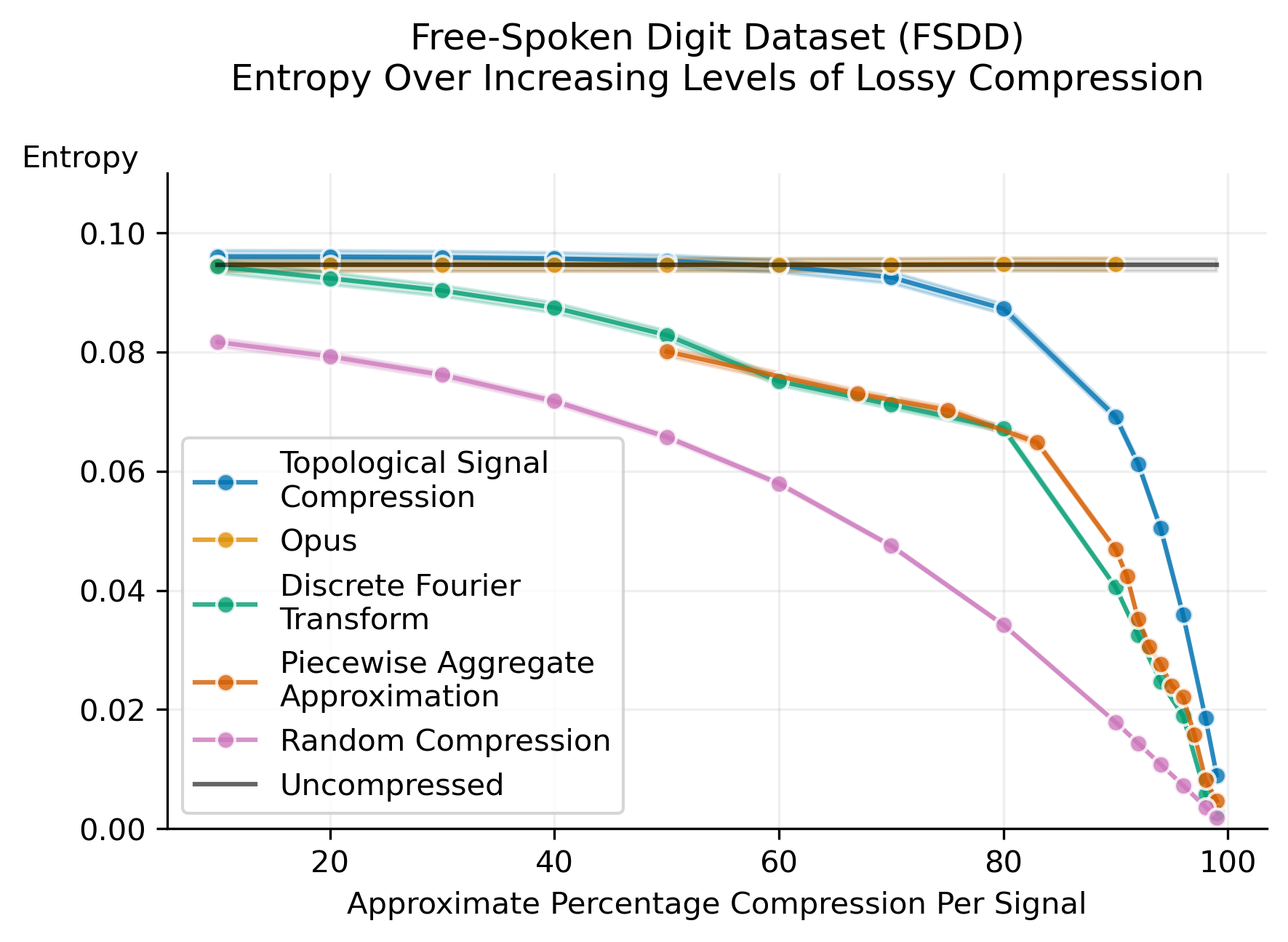}
	\caption{Entropy calculations on increasingly-compressed Free-Spoken Digit Dataset (FSDD) signals. Error bars represent 2 times the standard error of entropy over all digits in the dataset. TSC maintains entropy levels comparable to Opus at up to 70\% compression and similarly to Figure \ref{fig:ml_results} outperforms other compression methodologies above 90\% compression.}
	\label{fig:entropy}
\end{figure}

To abstract away from the potential variation in performance due to the machine learning training methodology, we also explored changes in entropy, which allowed us to measure the average amount of information lost over increasing levels of lossy compression for each method. We utilized the approximate entropy algorithm developed in \cite{Pincus2297} and expanded upon in \cite{ApEn} to measure average entropy across samples of FSDD for increasing compression levels, with the results shown in Figure \ref{fig:entropy}. Relative ``performance'' here is mostly consistent with machine learning classification accuracy, with the most notable exception being TSC maintaining comparable entropy to Opus at up to 70\% compression.

\subsection{Robustness to Noise}

To test these compression methodologies' capabilities on noisier data, we first mean-centered and standardized each signal in the dataset. We then added increasing levels of Gaussian noise to the standardized dataset. Finally, we compressed the noised signals with each compression methodology, after which we ran them through the featurization and machine learning pipeline the same as before. The resulting machine learning classification accuracies are shown in Figure \ref{fig:accuracy_with_noise}.

Although classification accuracy declines as more noise is added to the dataset, our results show consistent relative accuracy between the 5 compression methods for the unaltered signals and the noisy signals with up to 2.5 times the noise of the standardized signals added to the dataset, with one major exception being DFT compression performing noticeably worse with more noise. Although TSC's performance suffers more than Opus with increasing levels of noise, TSC still outperforms the other compression methodologies at greater than 90\% compression.

\section{Discussion}
\label{sec:discussion}

In addition to its superior machine learning performance at higher levels of compression with FSDD, Topological Signal Compression offers several additional benefits over the informed\footnote{The Random Compression scheme is highly interpretable, but uninformed by design as it compresses. With Random Compression serving mainly as a baseline control, it will be excluded from further discussion in this section.} competing compression methods.

\subsection{Interpretability When Changing the Compression Parameter}

We find TSC's underlying parameter for increasing compression, \textit{persistence}, to be more interpretable in its effect on the compressed signal than the other competing compression methodologies.

\begin{figure}[h!]
	\centering
	\includegraphics[width=3.4in]{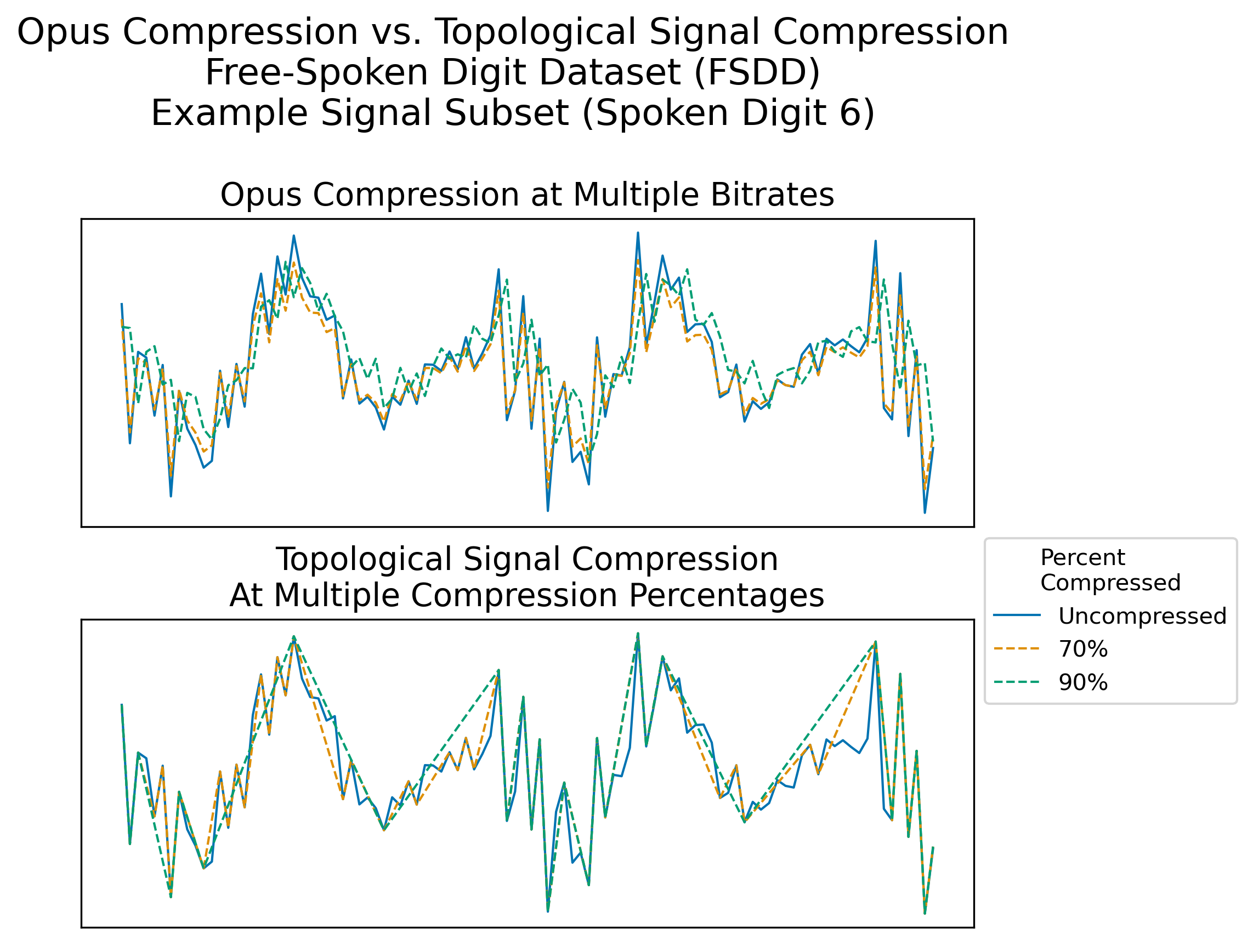}
	\caption{Comparing a reconstruction of the FSDD signal used in Figures \ref{fig:tsc_simplification_real_data} and \ref{fig:other_compressions_real_data}, zoomed in on a 1000 point subset of the signal. Note that at the 90\% compression level, TSC preserves the location of critical points while Opus does not.}
	\label{fig:tsc_vs_ogg}
\end{figure}

For Opus, changing the bitrate appears to have ambiguous effects on the $(t,f(t))$ pairs at higher levels of compression, though it does track well with the signal overall, as demonstrated in Figure \ref{fig:tsc_vs_ogg}. TSC, on the other hand, precisely recovers the critical points it keeps, which may be important for downstream use and interpretation of compressed signals.

\begin{figure*}
	\centering
	\includegraphics[width=7in]{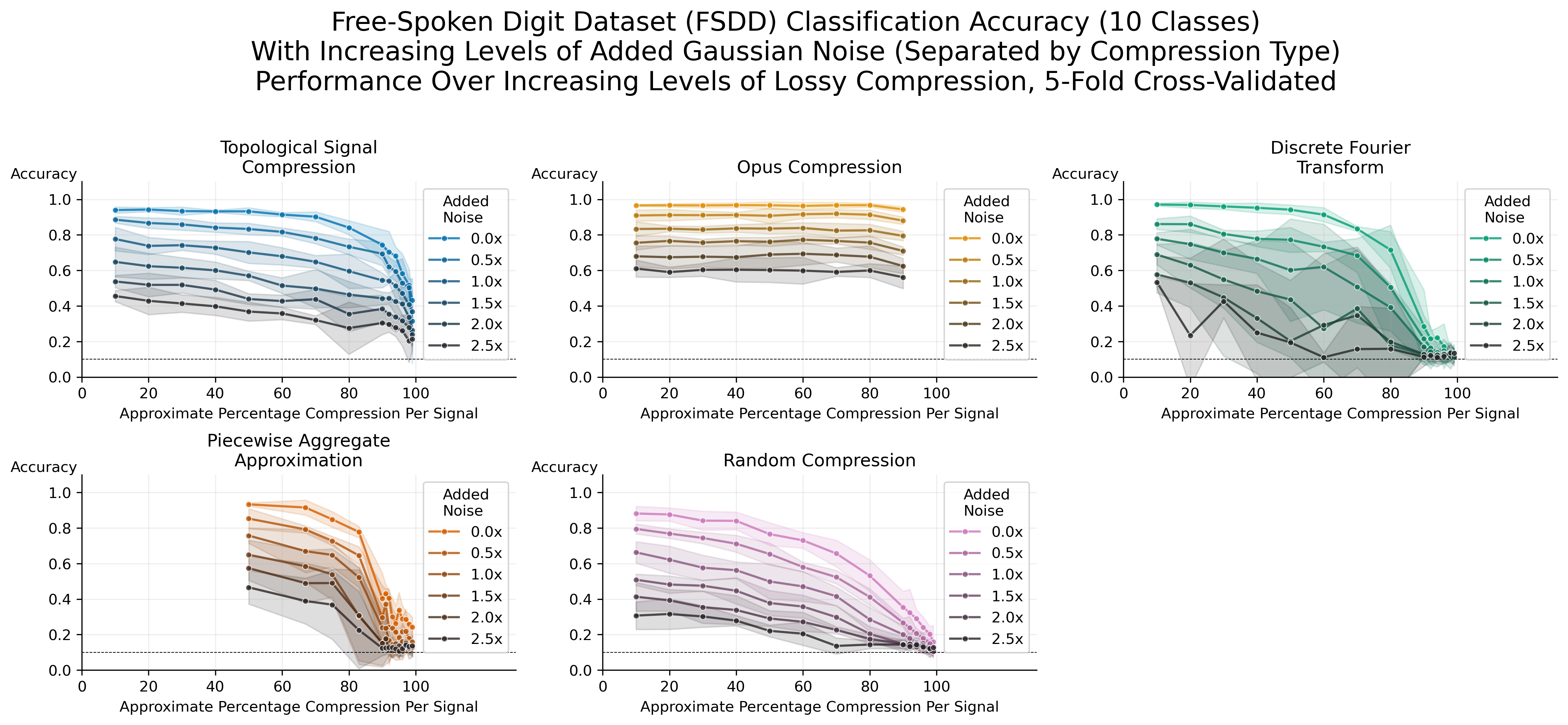}
	\caption{Accuracy of various compression methodologies over increasingly noisy FSDD dataset. Gaussian noise as high as 2.5 times the noise in the standardized signals was added to the dataset. Though machine learning accuracy declines as noise increases for all compression methodologies, TSC and Opus produce more consistent results, as exemplified by the smaller error bars and slower, smoother decline in machine learning accuracy at higher compression levels, even at high levels of noise. Note, we are visualizing the same data as in Figure \ref{fig:accuracy_with_noise}, instead separated by compression method as opposed to the amount of noise. Error bars represent 2 times the standard deviation of accuracy over the 5-fold cross-validated results at each compression level.}
	\label{fig:accuracy_with_noise_breakup_compression}
\end{figure*}

For PAA, summarizing a partitioning of windows over the signal sacrifices within-window distributional information. For example, skewed data in a window would result in taking a skewed mean value, and even if we take the median instead, we would remove the skew information content in the window. Furthermore, any repeating pattern risks washing out in a windowed summary statistic even at relatively low levels of compression. For example, a regularly sampled sine curve with window size of $2\pi$ would result in a compressed signal of only $0$ values. Although TSC will remove distributional information contained in non-critical points, it will otherwise preserve critical point distributional information. Additionally, repeating patterns will be preserved with TSC as long as they are sufficiently persistent.

For DFT, though it is excellent at preserving repeating patterns by its design, removing a Fourier coefficient does not have a strong intuitive implication for the resulting reconstructed signal, whereas removing a persistence pair with TSC has a localized, marginal  effect on the compression, discussed further in the next sub-section.

\subsection{Localized ``On-the-Margin'' Compression}

For all of our competing compression methods, tweaking the main parameter (bitrate for Opus, window size for PAA, and number of Fourier Coefficients for DFT) results in a \textit{global} change to the reconstructed signal. For TSC, however, we have an obvious means to make a \textit{marginal} change to the reconstructed signal. Starting from a given reconstruction, if we choose to remove only a single additional persistence pair from the reconstruction, we will ``un-kink'' a small subset of the reconstruction, leaving the remaining reconstruction unaltered. This results in increased interpretability \textit{between} reconstructions at variable compression levels, which supports the ability to use TSC in an environment with a variable communications budget, discussed in more detail later in this section.

\subsection{Handling of Noise}

The extent to which the compression algorithms perform well with respect to noise deserves special consideration since many real-world applications of compression occur in noisy environments. To explore this in the context of our machine learning exercises, we will reference Figure \ref{fig:accuracy_with_noise_breakup_compression}, which is simply the data from Figure \ref{fig:accuracy_with_noise} but separated instead by compression methodology to allow us to compare each method's performance against itself as noise increases.

DFT seems to be both theoretically and practically the least robust to noise. Theoretically, as noise increases, the risk of overfitting a Fourier coefficient to local noise within a signal and thus disrupting the signal globally increases. Furthermore, with potentially more poorly-formed Fourier coefficients, DFT should pay an additional penalty as one increases compression by keeping a smaller number of Fourier coefficients. In practice, this appears to hold true in Figure \ref{fig:accuracy_with_noise_breakup_compression}, with the error bars in machine learning accuracy spanning more than 40 percentage points at some compression levels as noise increases. We should note, however, an obvious means of improving DFT would be to run a ``windowed'' DFT, where to keep $n$ points, one would partition a signal into $m$ windows and keep $m/n$ points per window. Of course, windowing would also be a natural extension for TSC and even Random Compression as well, but we chose to only explore running these compression algorithms ``globally'' in this paper.

PAA should be theoretically robust to noise as long as the noise is unbiased. In particular, as window size increases, the chance of the noise in a window averaging out to $0$ increases. In practice, looking at Figure \ref{fig:accuracy_with_noise_breakup_compression}, this seems to hold, with the larger error bars at higher noise levels appearing to narrow as compression increases. Varying stability costs with respect to window size as noise increases is a clear negative to using PAA; however, it should be noted that PAA can likely play a strong role in situations with unbiased noise and very high sampling rates (where large windows can average out the noise without excessively compressing the actual information content in the signal).

The interpretability of noise on Opus compressions is somewhat uncertain, mainly due to the same concern raised by Figure \ref{fig:tsc_vs_ogg}, but the machine learning results were highly stable to noise over increased compression levels in practice, as demonstrated by the relatively small error bars for Opus in Figure \ref{fig:accuracy_with_noise_breakup_compression}.

TSC has a convenient interpretability when it comes to noise. Although the returned critical points may be shifted by noise, as soon as our persistence cutoff exceeds our noise level, the resulting reconstructed signals will be otherwise unaffected by the noise. This offers a generalizable, interpretable means by which one could insulate a model from minor amounts of noise. For example, if one had noiseless training data and were worried about the external validity of the resulting trained model to signals with small amounts of noise, then one could instead train and use the model in practice with TSC-simplified, noise-reduced signals, that is, signals with low-persistence critical points removed. As for TSC's machine learning results in practice, although showing a greater decline in classification accuracy when compared to Opus, Figure \ref{fig:accuracy_with_noise_breakup_compression} shows a stability in ML classification that is visually comparable to Opus, as exemplified by the relatively small error bars and smooth decay in accuracy for TSC both as noise increases and along the span of compression levels for a given amount of noise.

\subsection{Variable Communications Budget}

In order to make the most of a tight communications budget, a compression algorithm must have the flexibility to generate outputs of a precise size. For any given tight communications budget, there exists a \textit{nearly exact} compression level at which TSC could send a topologically simplified signal due to TSC's ability to throw out individual points from a given compression on the margin (e.g. sending one less critical point from the persistence diagram) as opposed to making global changes to the signal being compressed, as is done when changing compression levels using Opus and PAA (a marginal change to the bitrate or window size, respectively, will change the size of the compression more dramatically).

If the tight communications budget were \textit{variable}, then the marginal compression capability of TSC at the level of \textit{bytes} would allow one to efficiently utilize the entire communications budget by reconstructing the signal using the exact number of points allowed under the variable budget at any given moment in time.

In our earlier machine learning exercises, in particular Figures \ref{fig:ml_results} and \ref{fig:compressed_confus}, we trained a machine learning model for each specific level of compression. If one were sending compressed data over a variable communications budget, however, then one would also need to be ready to \textit{learn from variably-compressed data}. This would require a model (or set of models) that can be trusted over a range of compression levels.

Setting the valid compression range of a model poses a challenge. If the ranges get too wide, machine learning accuracy would likely suffer, but if the ranges are too narrow, one risks spreading the data too thin to reliably train models to span all ranges. Problems in either direction would reduce the trustworthiness of any modeled insights. The implied optimization scheme here is thus to maintain the largest possible compression ranges as long as the resulting signals are sufficiently comparable to deserve being classified by the same model. Therefore, the more smoothly the signals change as they are compressed, the greater the range of compression levels one would expect to be able to reliably utilize a given trained model.

In addressing this concern, we will focus only on TSC and DFT, as they are the only two informed algorithms considered in this paper that can easily achieve a byte-specific variable communications budget.

\begin{figure*}[h!]
	\centering
	\includegraphics[width=3.4in]{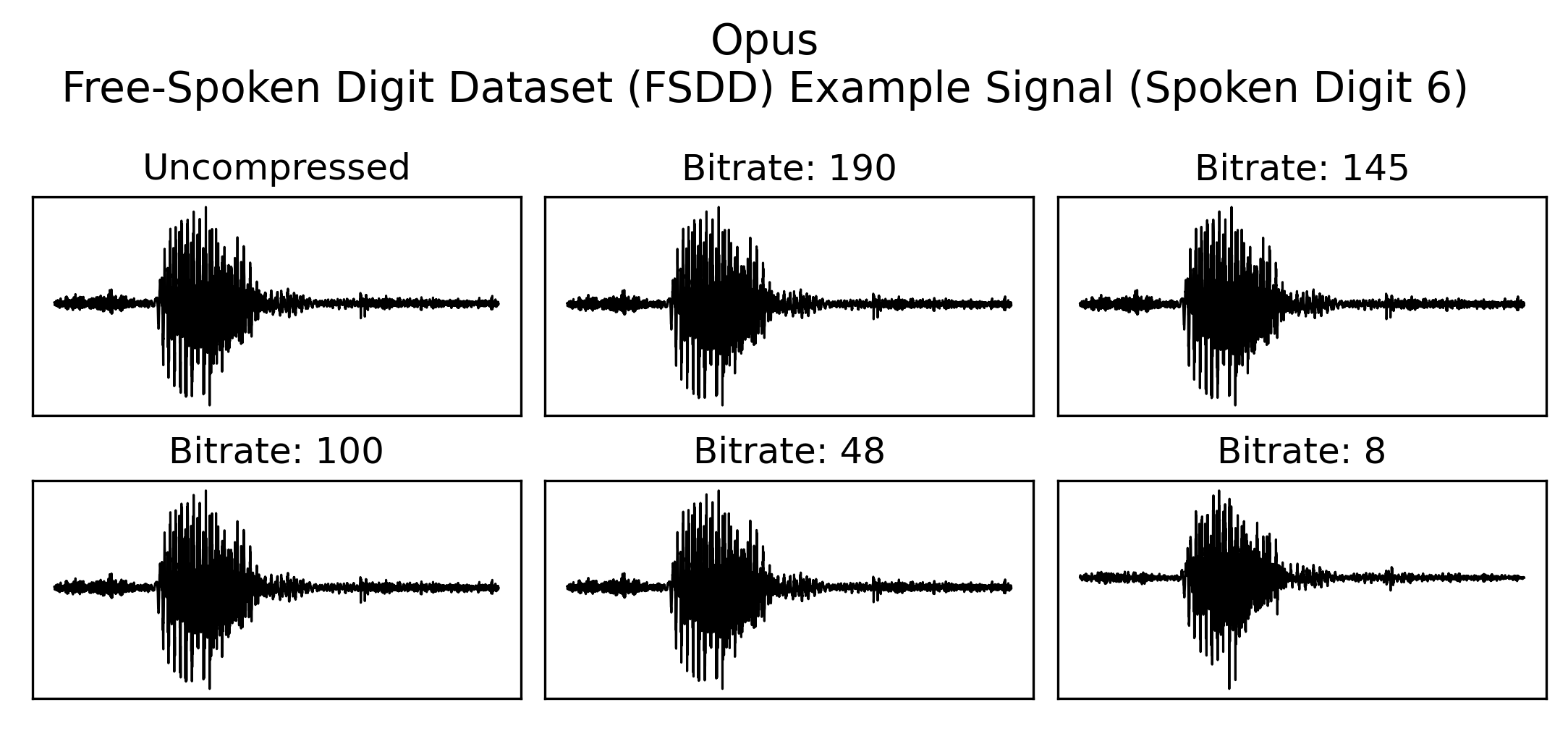}\quad\includegraphics[width=3.4in]{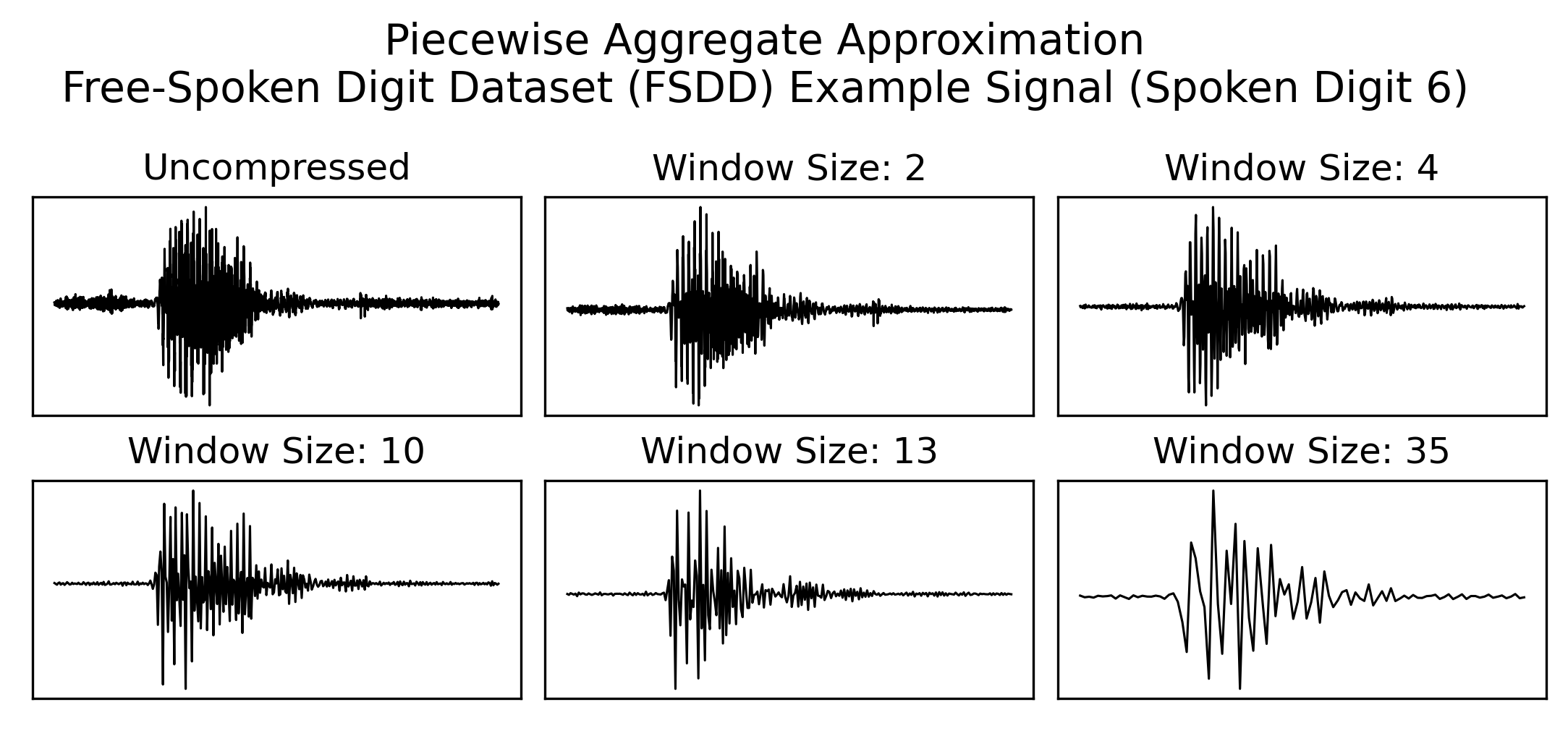}
	\includegraphics[width=3.4in]{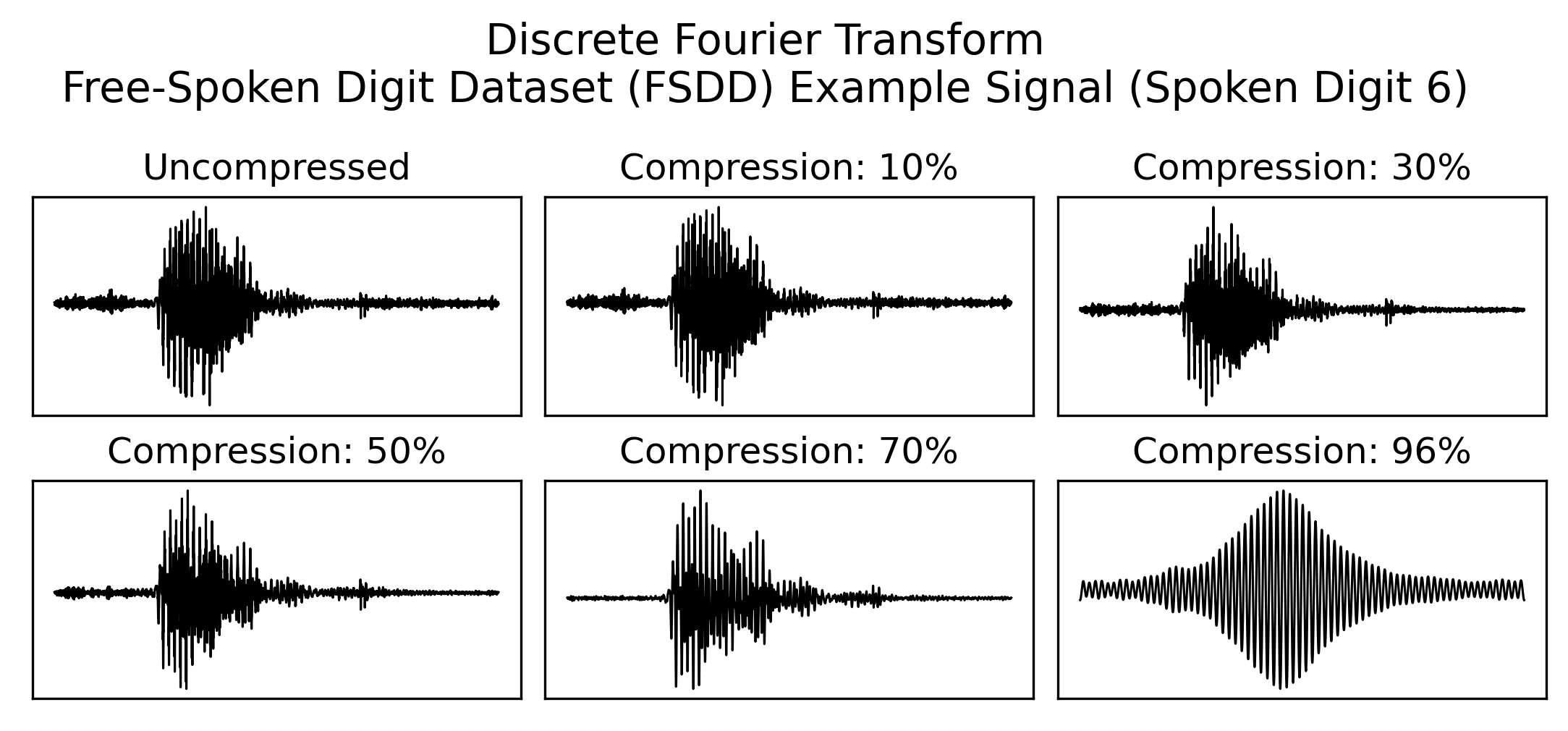}\quad\includegraphics[width=3.4in]{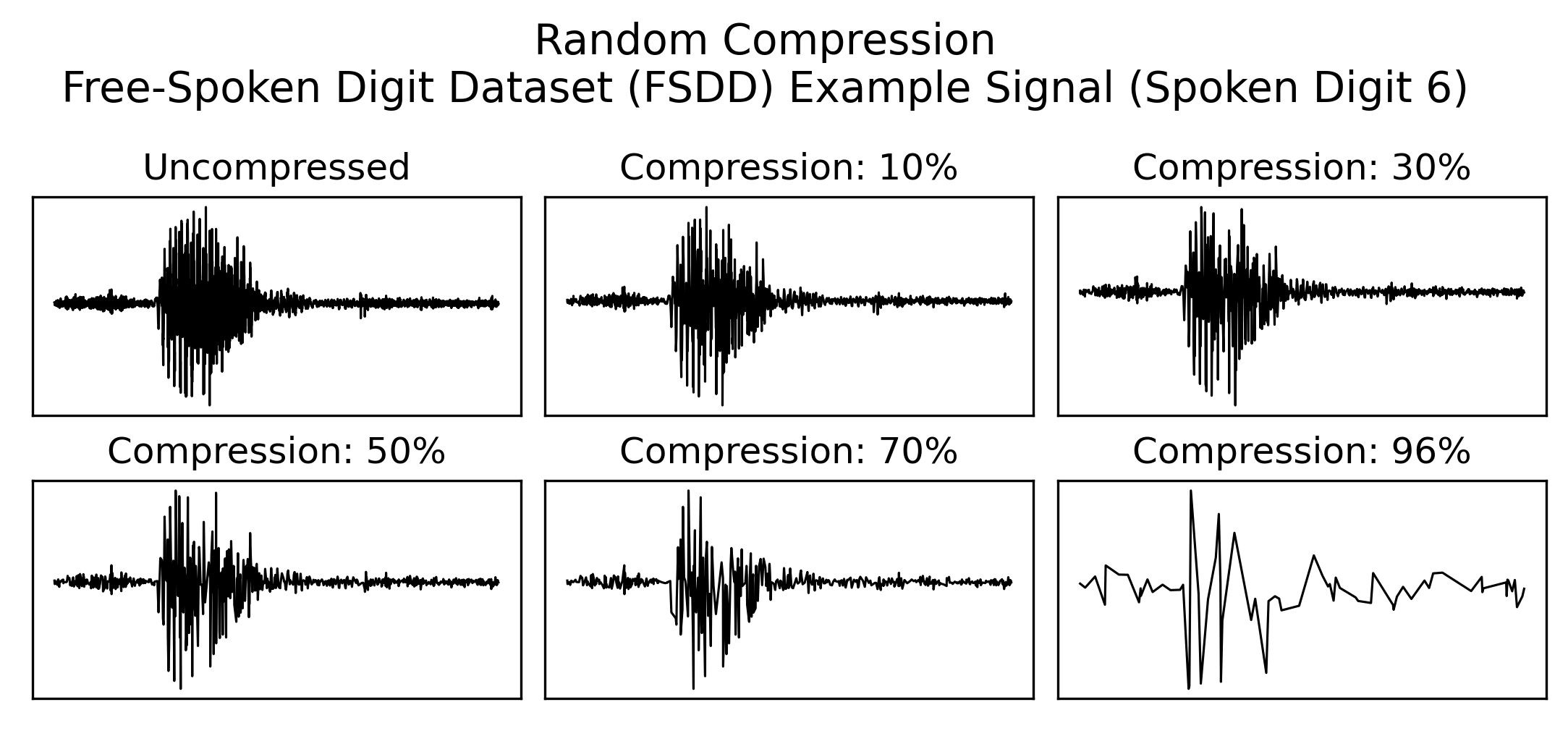}
	\caption{Counterfactual compression methodologies run on the same FSDD signal as in Figure \ref{fig:tsc_simplification_real_data}. Note: the bitrate compressions for Opus and window size compressions for Piecewise Aggregate Approximation do not correspond to the same compression percentages shown for Topological Signal Compression, Random Compression, and the Discrete Fourier Transform.}
	\label{fig:other_compressions_real_data}
\end{figure*}

On a theoretical level, we should already expect TSC to excel at this task relative to DFT, based on the earlier discussion of TSC's ``marginal'' compression effects on a signal as opposed to DFT's ``global'' compression effects. On an anecdotal level, note in Figures \ref{fig:tsc_simplification_real_data} and \ref{fig:other_compressions_real_data}, the ``shape'' of the compressed signal is relatively consistent for TSC but changes drastically for DFT at higher levels of compression.

\begin{figure}[h!]
	\centering
	\includegraphics[width=3.4in]{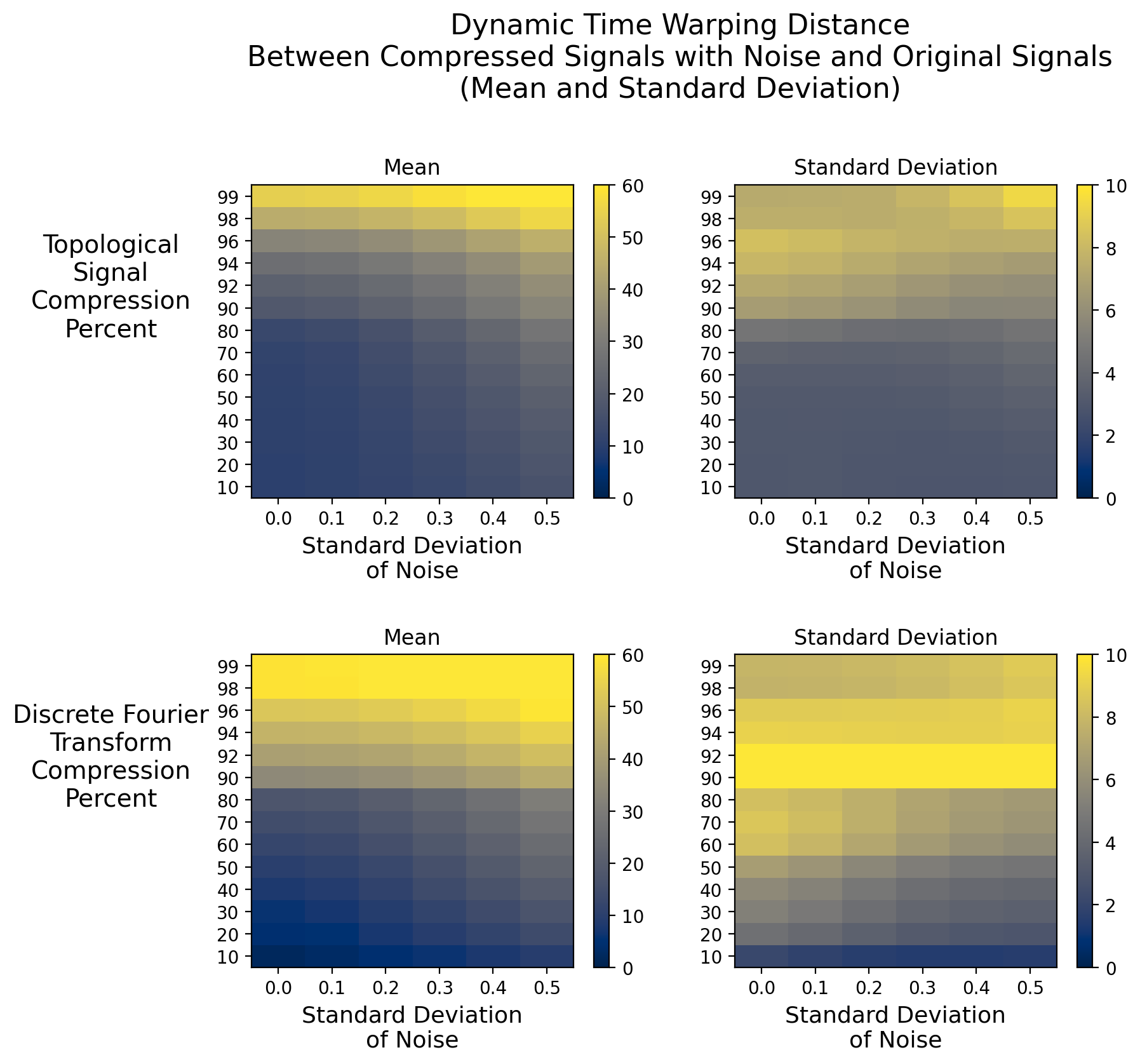}
	\caption{Dynamic Time Warping (DTW) distance between original signals and reconstructions of the FSDD signals compressed using Topological Signal Compression (TSC) and Discrete Fourier Transform (DFT). Although DFT achieves closer DTW distance and lower variance of distances at lower levels of compression, TSC outperforms DFT on both fronts at higher levels of compression. TSC also exhibits a smoother increasing of DTW distances than DFT.}
	\label{fig:dtw_distances}
\end{figure}

To empirically explore this, we looked at the Dynamic Time Warping distance between original and compressed signals over increasing amounts of both compression and Gaussian noise, shown in Figure \ref{fig:dtw_distances}. Despite DFT achieving greater similarity and lower variance at low levels of compression, TSC outperforms DFT on both fronts at higher levels of compression while additionally maintaining a more stable change both over increased compression and noise. Thus, the relative ``smoothness'' of Dynamic Time Warping distances for TSC over DFT is indicative of a greater ability for a TSC-trained model to generalize to a larger range of compression levels.

\section{Conclusion}
\label{sec:conclusion}

We find Topological Signal Compression offers the most promising ability to create actionable information under a variable communications budget. More generally, TSC's interpretability combined with its ability to fine-tune its compression to arbitrarily high byte constraints, locally compress on the margin, smoothly change its increasingly compressed signals, handle noise, and generalize to any signal data make it worthy of consideration in any constrained communications scenario.

\section*{Acknowledgments}
Research on this project was partially supported by the DARPA Ocean of Things project, under contract N6600121C4006.
This document has been Approved for Public Release, with Distribution Statement A, Distribution Unlimited.
The views, opinions and/or findings expressed are those of the authors and should not be interpreted as representing the official views or policies of the Department of Defense or the U.S. Government.

\bibliography{bibliography}
\bibliographystyle{ieeetr}

\end{document}